\documentclass[12pt]{article}
\usepackage{epsfig}
\usepackage{pst-plot,url,epsf}
\newcommand{\beq}{\begin{equation}}
\newcommand{\eeq}{\end{equation}}
\newcommand{\bea}{\begin{eqnarray}}
\newcommand{\eea}{\end{eqnarray}}
\newcommand{\nobody}{\rule{0ex}{1ex}}

\newcommand{\epm}{e^+e^-}

\newcommand{\ra}{\rightarrow}

\newcommand{\ttbar}{t\bar{t}}

\newcommand{\sixf}{b f_1 \bar{f'_1} \bar{b} f_2 \bar{f'_2}}
\newcommand{\eesixf}{e^+ e^- \ra b f_1 \bar{f'_1} \bar{b} f_2 \bar{f'_2}}
\newcommand{\eebnmbdu}{e^+ e^- \ra b \nu_{\mu} \mu^+ \bar{b} d \bar{u}}

\newcommand{\AmS}{{\protect\the\textfont2
  A\kern-.1667em\lower.5ex\hbox{M}\kern-.125emS}}

\newcommand{\eebnmbmn}{e^+ e^- \ra b \nu_{\mu} \mu^+ \bar{b} \mu^- 
                                                          \bar{\nu}_{\mu}}

\begin{document}
\thispagestyle{empty}
\begin{flushright}
April 2003\\
\vspace*{1.5cm}
\end{flushright}
\begin{center}
{\LARGE\bf Studying top quark pair production\\
  at a linear collider with a program \\[2mm]
      {\tt eett6f}}\footnote{Presented 
      at the Cracow Epiphany Conference on Heavy Flavors,
      Cracow, Poland, January 3--6, 2003.}$\nobody^,$\footnote{Work supported
           in part by the Polish State Committee for Scientific Research
           (KBN) under contract No. 2~P03B~045~23 and by European 
  Commission's 5-th Framework contract  HPRN-CT-2000-00149.}\\
\vspace*{2cm}
Karol Ko\l odziej\footnote{E-mail: kolodzie@us.edu.pl}\\[1cm]
{\small\it
Institute of Physics, University of Silesia\\ 
ul. Uniwersytecka 4, PL-40007 Katowice, Poland}\\
\vspace*{3.5cm}
{\bf Abstract}\\
\end{center}
Some features of a program {\tt eett6f} for top quark pair production and 
decay into six fermions at linear colliders are discussed. 
Lowest order standard model predictions for cross sections of
some six fermion channels and for the top quark decay
width are confronted with the predictions obtained within a model with
an anomalous $Wtb$ coupling.
The question of wether non doubly resonant background can easily be
reduced by imposing kinematical cuts is addressed.

\vfill
\newpage

\section{INTRODUCTION}

As a top quark is the heaviest matter particle ever observed, 
with mass close to the energy scale of the electroweak symmetry breaking,
investigation of its physical properties may give hints towards
understanding physics beyond the Stnadard Model at higher energy scales.
Therefore precise measurements, at a level of a few per mille, of the top 
quark properties and interactions will most certainly belong to the research 
program of any future $\epm$ linear collider \cite{NLC}.
In order to match that high precision level of the measurements, theoretical 
predictions should include radiative corrections.

As it has been shown in \cite{KK} and will be illustrated later in this 
lecture, effects caused by the off-shellness 
of the $\bar{t}t$ pair and off-resonance contributions may be
important too, especially for measurements at the centre of mass system 
(c.m.s.) energies much above the $t\bar{t}$ threshold. As the $t$-
and $\bar{t}$-quark of the reaction 
\beq
\label{eett}
         \epm \ra \bar{t} t
\eeq
almost immediately decay, predominantly into $bW^+$ 
and $\bar{b}W^-$, with the $W^{\pm}$ boson decaying into a fermion pair, 
what one actually observes are 6 fermion reactions of the form
\bea
\label{eesixf}
         \eesixf,
\eea
where $f_1=\nu_e, \nu_{\mu}, \nu_{\tau}, u, c$, $f_2=e^-, \mu^-, \tau^-, d, s$ 
and $f'_1$, $f'_2$ are the corresponding weak isospin partners,
$f'_1=e^-, \mu^-, \tau^-, d, s$, $f'_2=\nu_e, \nu_{\mu}, \nu_{\tau}, u, c$.
For example, a pure electroweak (EW) reaction 
\bea
\label{eebnmbmn}
                 \eebnmbmn
\eea
 in unitary gauge receives contributions from 452 Feynman diagrams,
neglecting the Higgs boson couplings to fermions lighter than $b$,
while there are only two `signal' diagrams contributing 
to reaction (\ref{eesixf}) in the double resonance approximation
\bea
\label{eettsixf}
         \epm \ra \bar{t}^* t^* \ra \sixf
\eea
at the same time.

In the present lecture, some features of the updated version of a computer 
program {\tt eett6f}  \cite{KK} that allows for computing cross sections 
of reactions (\ref{eesixf}) to the lowest order of the standard model (SM), 
with a complete set of the Feynman diagrams, will be discussed.
The question of wether the non doubly resonant background can easily be
reduced by imposing kinematical cuts will be addressed.
Moreover, lowest order SM predictions for some six fermion 
channels of (\ref{eesixf}) and for the 3 particle top quark decay
width will be confronted with the predictions obtained within a model with
an anomalous $Wtb$ coupling. 

\section{A PROGRAM}

A computer program {\tt eett6f} for calculating cross sections of 
6 fermion reactions (\ref{eesixf}) relevant for a $\ttbar$-pair production 
and decay at c.m.s. energies typical for linear colliders has been written
in {\tt FORTRAN 90} \cite{program}. The program consists of 50 files 
including a makefile, 
all stored in one working directory. The user should 
specify the physical input parameters in a module {\tt inprms.f} and 
select a number of options in the main program {\tt csee6f.f}. 
The options allow, among other, for calculation
of the cross sections while switching on and off different subsets of the 
Feynman diagrams.
It is also possible to calculate cross sections in two different
narrow width approximations, for $\ttbar$-quarks, or $W^{\pm}$-bosons.
The program allows for taking into account both the EW
and QCD lowest order contributions.
Version~1 of the program allows for calculating both the 
total and differential cross sections of (\ref{eesixf}) at tree level of SM. 
Some anomalous effects beyond the SM that will be discussed later
have been already implemented in the program. The program can be used as the 
Monte Carlo (MC) generator of unweighted events as well.

Matrix elements are calculated with the helicity amplitude method 
described in \cite{KZJ}.
Phase space integrations are performed with the MC method.
The most relevant peaks of the matrix element squared, related to 
the Breit-Wigner shape of the $W, Z$, Higgs and top quark resonances,
and to the exchange of a massless photon, or gluon
have to be mapped away. As it is not possible to find out a single 
parametrization of the 
$14$-dimensional phase space which would allow to cover the whole 
resonance structure of the integrand, the program utilizes
a multichannel MC approach.

Constant widths of unstable particles: the massive electroweak vector 
bosons, the Higgs boson and the top quark, are introduced through the complex 
mass parameters
\bea
\label{masses}
 M_V^2 &\!\!=\!\!& m_V^2-im_V\Gamma_V, \quad V=W, Z, \\
 M_H^2&\!\!=\!\!& m_H^2-im_H\Gamma_H,  
                         \quad M_t=m_t-i\Gamma_t/2, \nonumber
\eea
which replace masses in the corresponding propagators, 
\bea
\label{props}
\Delta_F^{\mu\nu}(q)\!&=&\!\frac{-g^{\mu\nu}+q^{\mu}q^{\nu}/ M_V^2}
                               {q^2- M_V^2},   \\
\Delta_F(q)\!&=&\!\frac{1}{q^2- M_H^2}, \qquad
S_F(q)=\frac{/\!\!\!q+ M_t}{q^2- M_t^2}. \nonumber 
\eea
Propagators of a photon and a gluon are taken in the Feynman gauge.

The EW mixing parameter may be defined either real or complex,
\bea
\label{sintw}
\sin^2\theta_W=1-\frac{m_W^2}{m_Z^2}, \quad {\rm or} \quad
\sin^2\theta_W=1-\frac{M_W^2}{M_Z^2}.
\eea
Performing substitution (\ref{props}) both in 
the $s$- and $t$-channel Feynman diagrams and taking the complex electroweak
mixing parameter of Eq.~(\ref{sintw}) leads to fulfilment of the
Ward identities. 
As light fermion masses are not neglected, cross sections
can be calculated without any kinematical cuts.

A number of checks of the program {\tt eett6f} have been performed.
The reader is referred to \cite{program} for details.

In the present version of the program, an anomalous $Wtb$ coupling has been
implemented. The model describing departures of the $Wtb$ 
interaction from the SM, caused by new fundamental 
interactions at high energies, can be specified in terms of the low 
energy effective lagrangian containing terms of dimension 6 \cite{kbg} 
\begin{eqnarray}
\label{lagr}
L = \frac{g}{\sqrt{2}} & & \hspace*{-3.5mm}
     \left[\; W_{\mu}^-{\bar b}\gamma^{\mu}
      \left(f_1^-P_{-}+f_1^+P_{+}\right)t \right. \nonumber \\
& &- \left. \frac{1}{m_{W}} \partial_{\nu} W_{\mu}^- {\bar b}\sigma^{\mu\nu}
      \left(f_2^-P_{-} + f_2^+P_{+}\right) t\right] + {\rm h.c.}
\end{eqnarray}
In Eq.~(\ref{lagr}), $P_{\pm}$ are chirality projectors 
given by $P_{\pm}=\left(1 \pm \gamma_{5}\right)/2$,
$\sigma^{\mu\nu}=i\left(\gamma^{\mu}\gamma^{\nu}-
\gamma^{\nu}\gamma^{\mu}\right)/2$, 
the anomalous couplings $f_1^{\pm}$ and $f_2^{\pm}$ are assumed to be real,
and $W^-$ should be regarded as
an effective vector field. 
The SM $Wtb$ coupling is reproduced with $f_1^+=f_2^-=f_2^+=0$
and $f_1^-$ equal to the real Cabibbo-Kobayashi-Maskawa 
matrix element $V_{tb}$.
According to the present experimental data, $V_{tb}$ is very close to 1
\cite{PDG} and $f_1^+$ is strongly constrained by the CLEO 
$b \rightarrow s \gamma$ data which give $f_1^+ \approx 0$ \cite{CLEO}.
The couplings $f_2^-$ and $f_2^+$ are at present only weakly constrained,
see {\em e.g.} \cite{whisnant}. 

\section{NUMERICAL RESULTS}

In the present section, a sample of numerical results obtained with
the current version of {\tt eett6f} will be presented. 
The physical input parameters that are used are the same as in \cite{KK}. 

The energy dependence of the total cross sections of (\ref{eebnmbmn}) is 
shown in Figure~1. The full lowest order cross section $\sigma$ 
is plotted with the solid line, the signal cross section 
$\sigma_{\bar{t}^*t^*}$ with the dotted line, and the cross section in 
the narrow top width approximation, $\sigma_{\bar{t}t}$, with the dashed line.
A comparison of the solid and dotted lines shows the effects of the
off resonance background contributions to reaction (\ref{eebnmbmn}),
and a comparison of the dotted and dashed lines shows the effect
of the off-mass-shell production of the $\ttbar$-pair. Both effects are
substantial, especially at higher energies. Therefore they should be
taken into account in the future analysis of data.


\begin{figure}[ht]
 \label{fig1}
\begin{picture}(35,35)(-60,-75)
\rput(5.3,-6){\scalebox{0.7 0.7}{\epsfbox{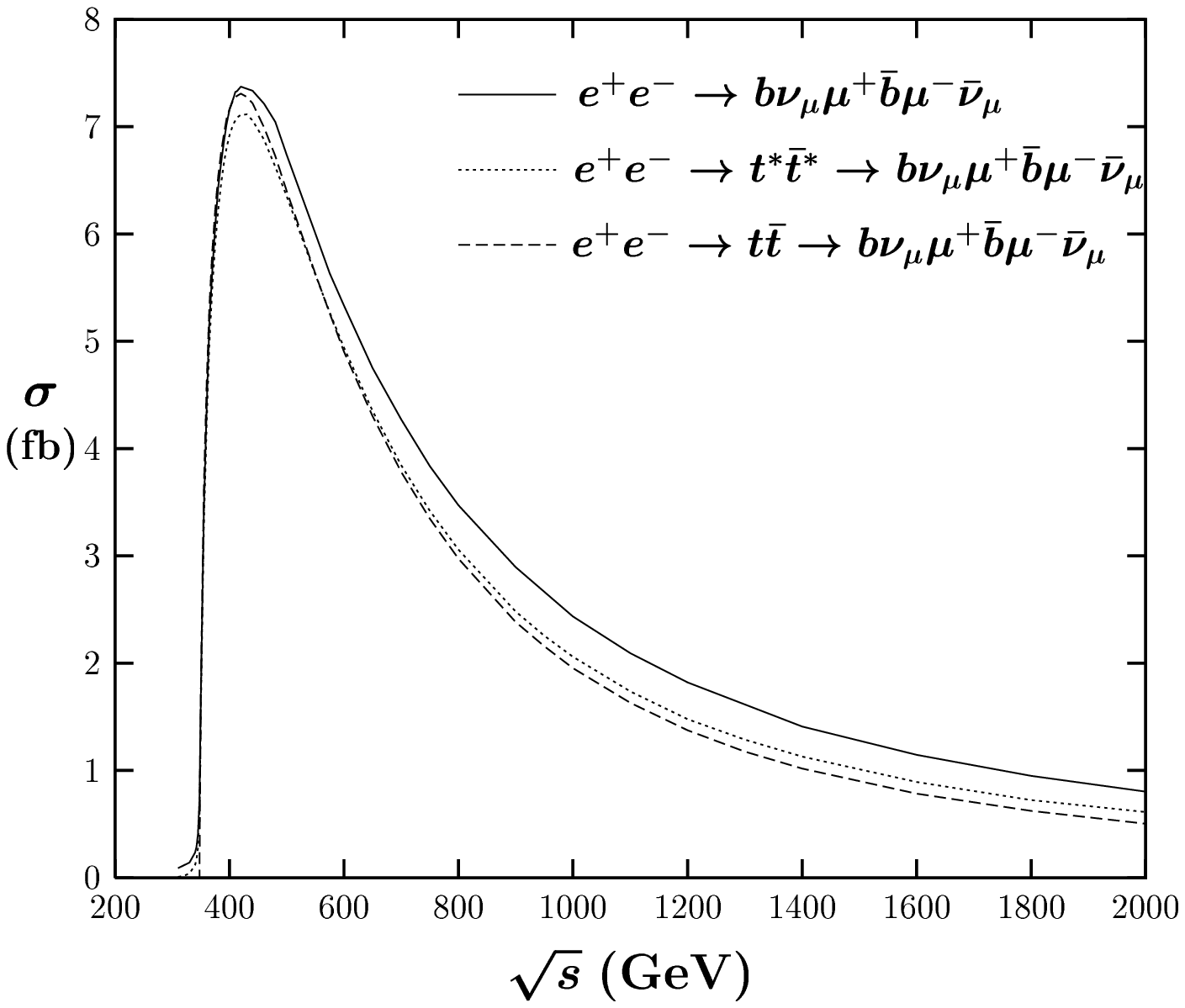}}}
\end{picture}
\vspace*{7.3 cm}
\caption{Total cross sections of $\eebnmbmn$ as functions of the 
          c.m.s. energy.}
\end{figure}

The question arises, whether those big off resonance effects and background
contributions can easily be reduced by imposing cuts. 
This issue will be preliminarily addressed below 
for one specific channel of reaction (\ref{eesixf})
by imposing the following
cuts that are usually used in literature
\bea
\label{cuts}
\begin{array}[b]{rlrl}
\theta (l,\mathrm{beam})> & 5^\circ, & \qquad
\theta (q,\mathrm{beam})> & 5^\circ, \\
\theta( l, l')> & 5^\circ, &
\theta (l, q)> & 5^\circ,\\
E_l > & 10\;{\rm GeV}, & E_q> & 10\;{\rm GeV},\\
 m(q,q')> & 10\;{\rm GeV}, &  
\end{array}  
\eea
where $q$, $l$ denote quark and charged lepton, respectively.

Results for three differential cross sections:
invariant mass distribution of a $\bar{b}d\bar{u}$-quark
triple and energy distributions of a $b$-quark and $\mu^+$
of $\eebnmbdu$ at the c.m.s. energy of 500 GeV with cuts (\ref{cuts})
have been compared against the corresponding results without cuts
in Figs.~2--4. The solid histograms in Figs.~2--4 show the lowest
order SM results obtained with the complete set of the Feynman
diagrams, the dotted histograms show the contribution of the
two $\ttbar$ signal diagrams and the dashed histograms have been obtained
in the narrow top width approximation. The dotted histograms in Fig.~2
cannot be distinguished from the solid ones.

\begin{figure}[ht]
\label{fig2}
\begin{center}
\setlength{\unitlength}{1mm}
\begin{picture}(35,35)(55,-50)
\rput(5.3,-6){\scalebox{0.6 0.6}{\epsfbox{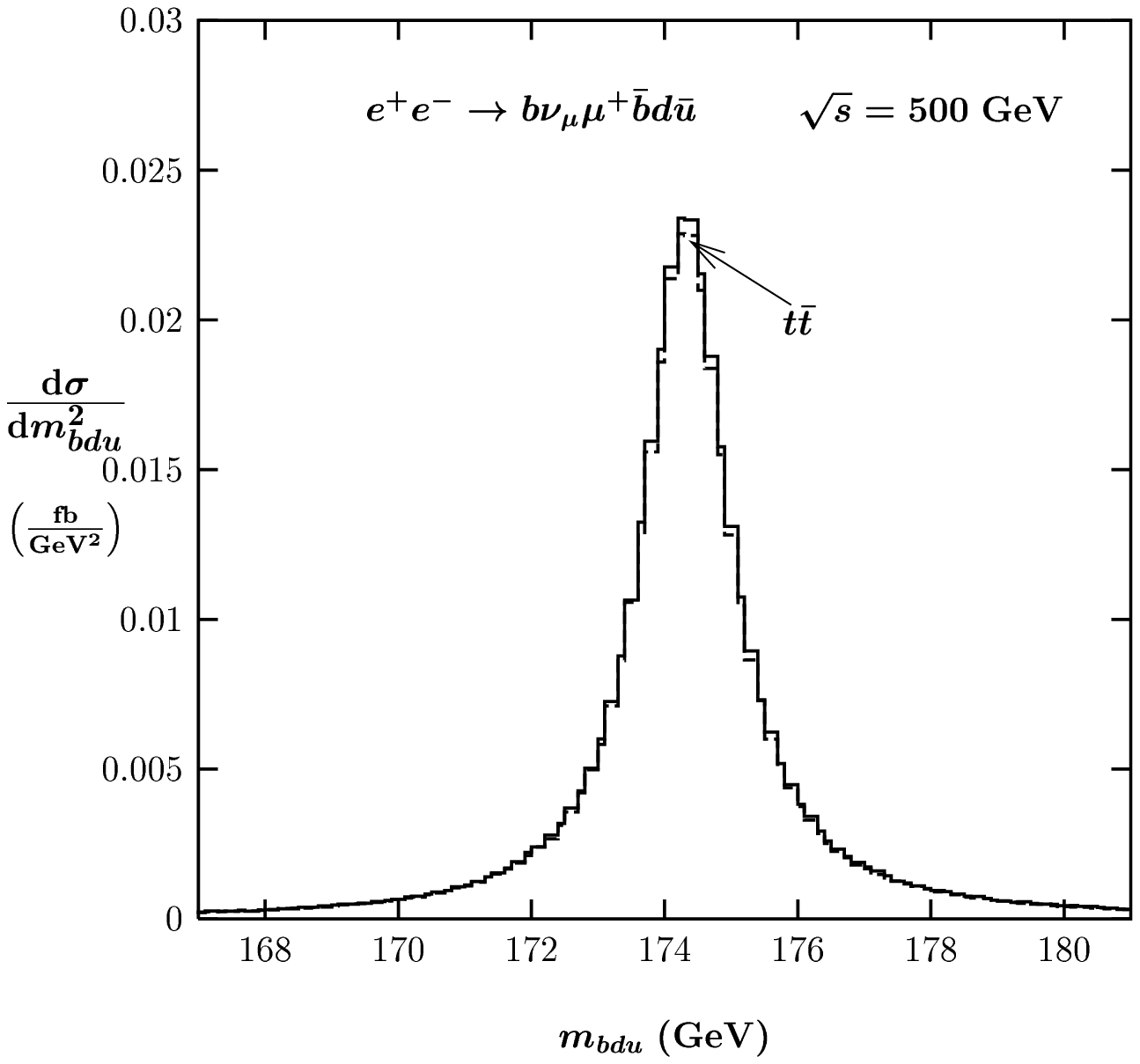}}}
\end{picture}
\begin{picture}(35,35)(15,-50)
\rput(5.3,-6){\scalebox{0.6 0.6}{\epsfbox{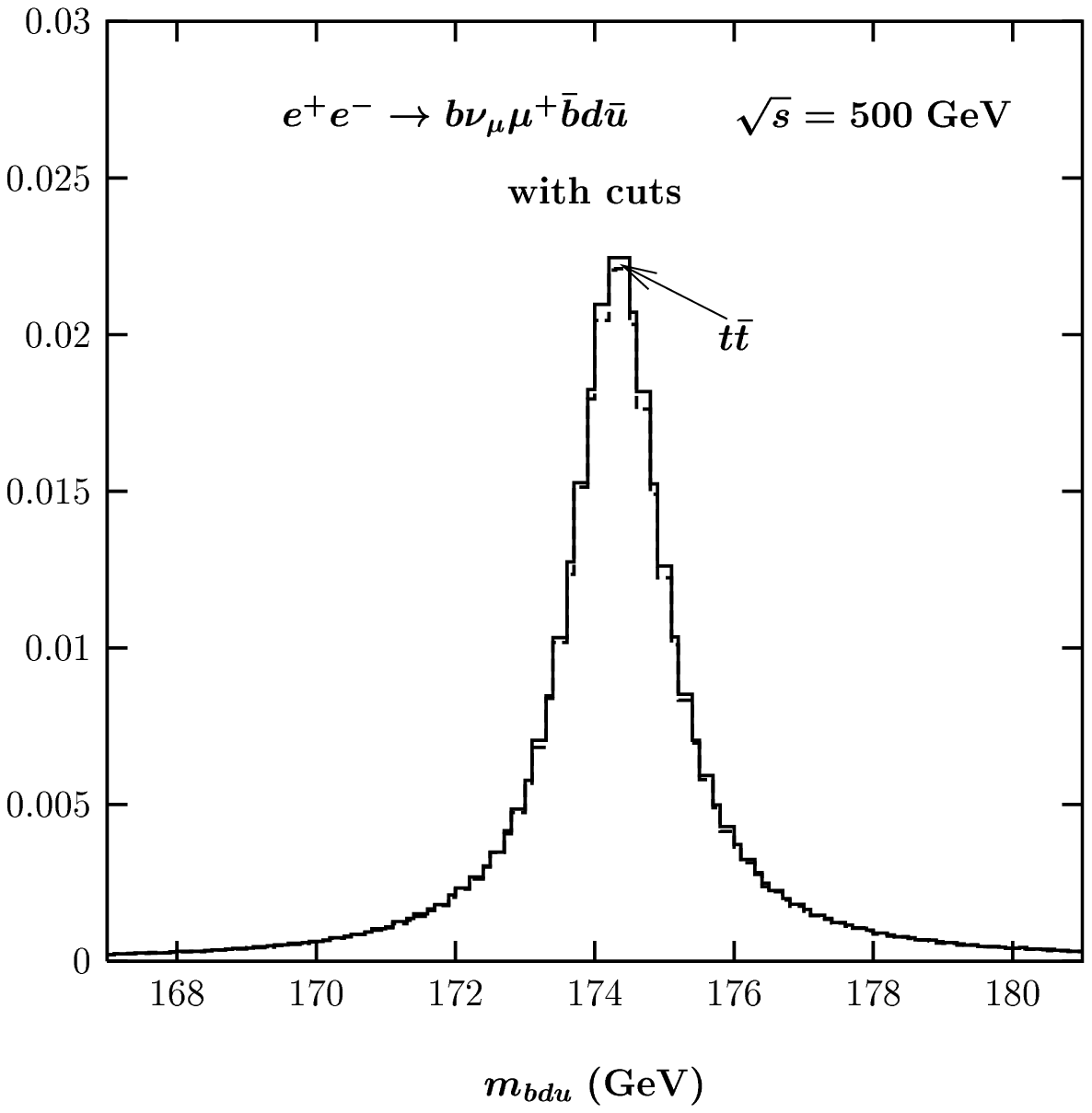}}}
\end{picture}
\end{center}
\vspace*{3.5cm}
\caption{Invariant mass distributions of a $\bar{b}d\bar{u}$-quark
triple at $\sqrt{s}=500$ GeV without cuts (left) and with cuts (right).}
\end{figure}


\begin{figure}[ht]
\label{fig3}
\begin{center}
\setlength{\unitlength}{1mm}
\begin{picture}(35,35)(55,-50)
\rput(5.3,-6){\scalebox{0.6 0.6}{\epsfbox{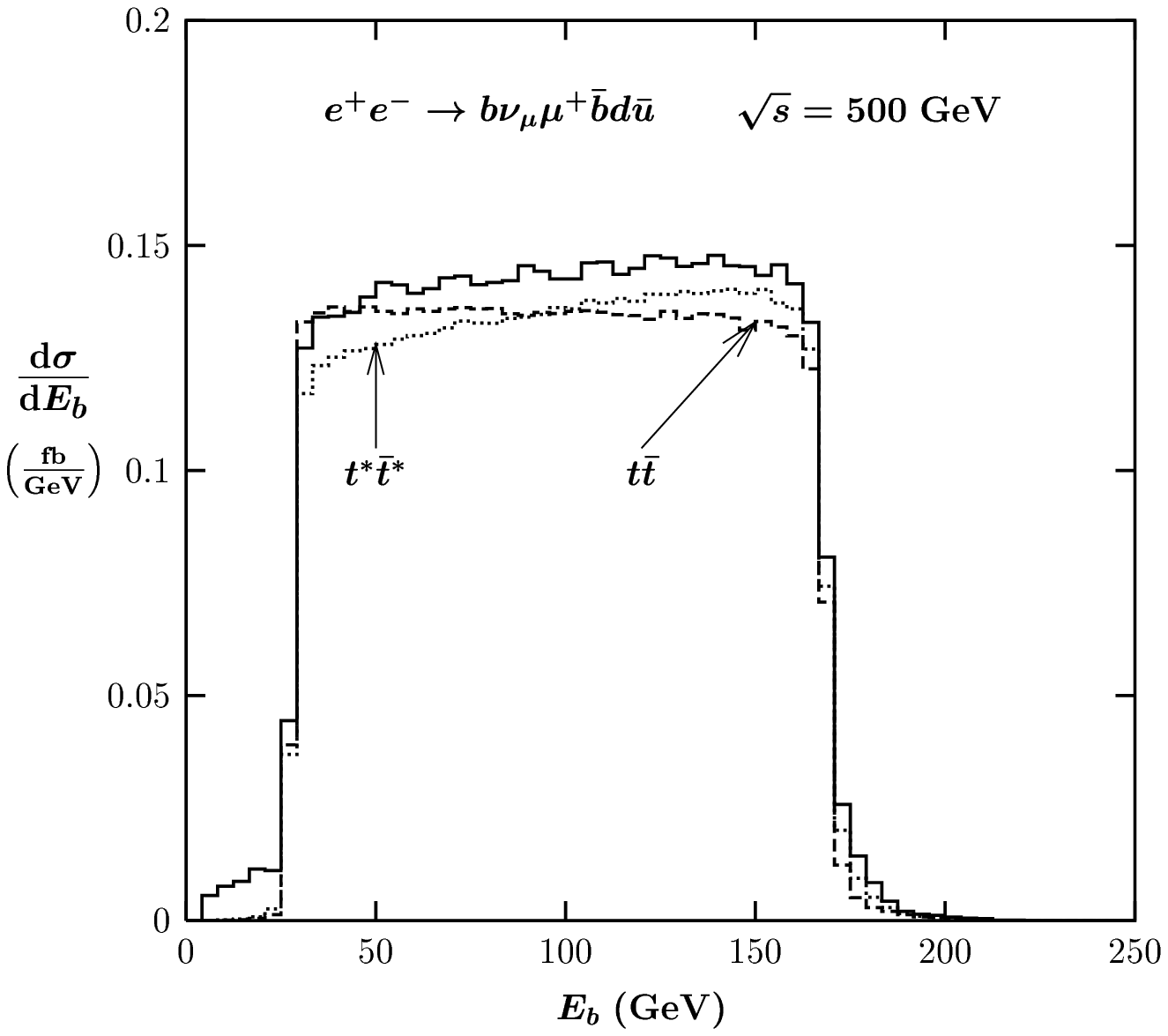}}}
\end{picture}
\begin{picture}(35,35)(15,-50)
\rput(5.3,-6){\scalebox{0.6 0.6}{\epsfbox{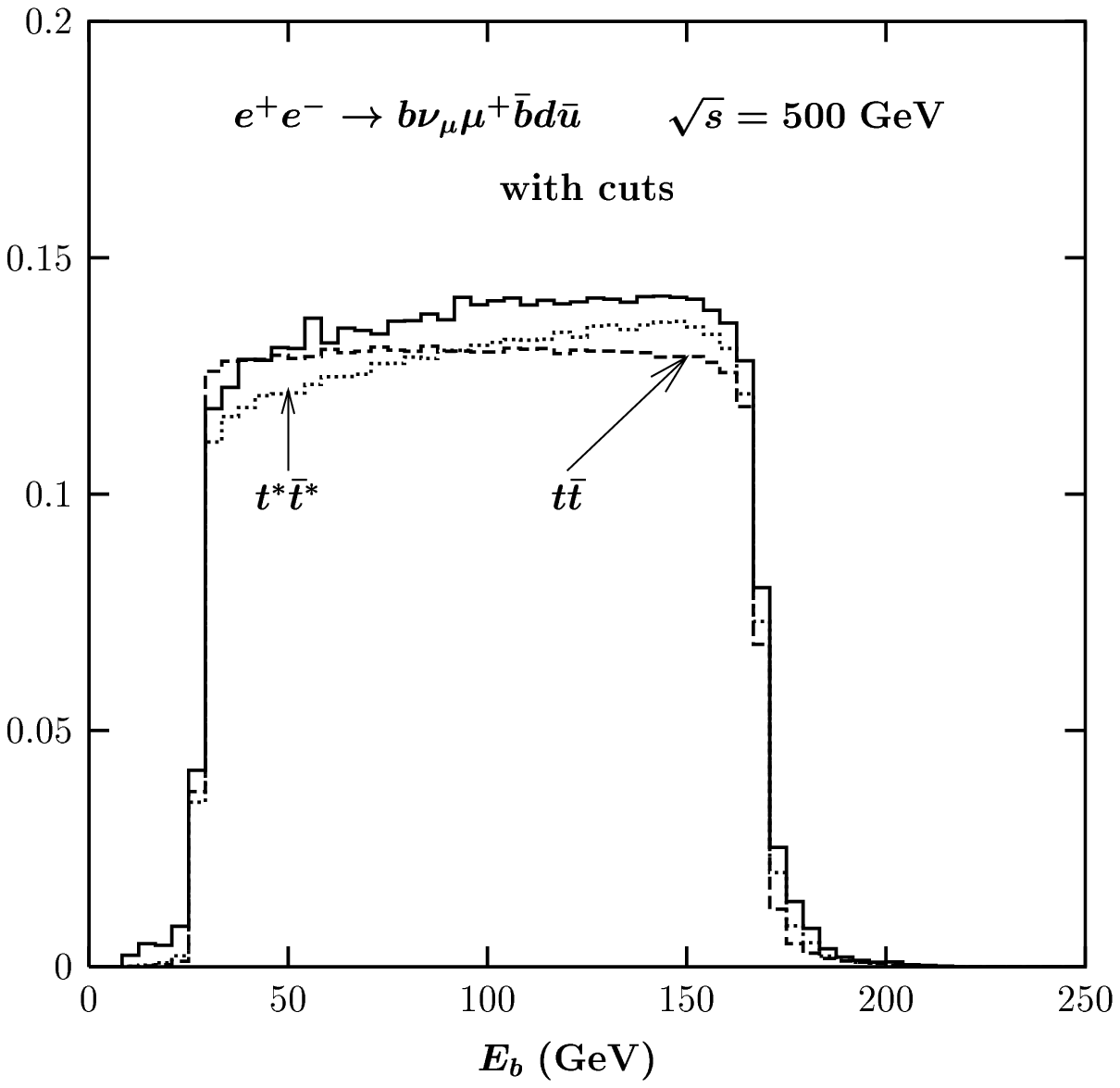}}}
\end{picture}
\end{center}
\vspace*{3.5cm}
\caption{Energy distributions of a $b$-quark at $\sqrt{s}=500$ GeV without cuts
and with cuts.}
\end{figure}

\begin{figure}[ht]
\label{fig4}
\begin{center}
\setlength{\unitlength}{1mm}
\begin{picture}(35,35)(55,-50)
\rput(5.3,-6){\scalebox{0.6 0.6}{\epsfbox{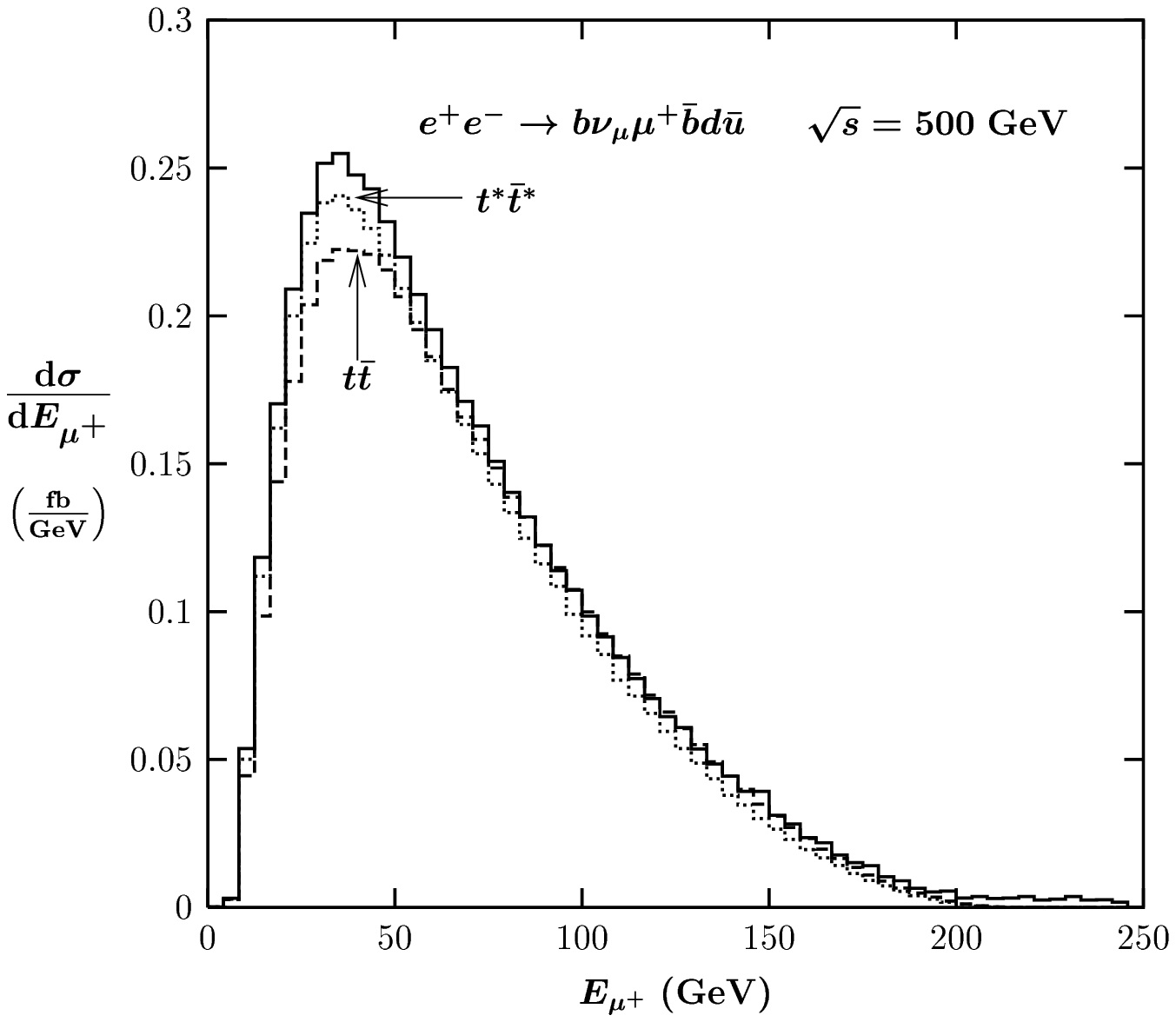}}}
\end{picture}
\begin{picture}(35,35)(15,-50)
\rput(5.3,-6){\scalebox{0.6 0.6}{\epsfbox{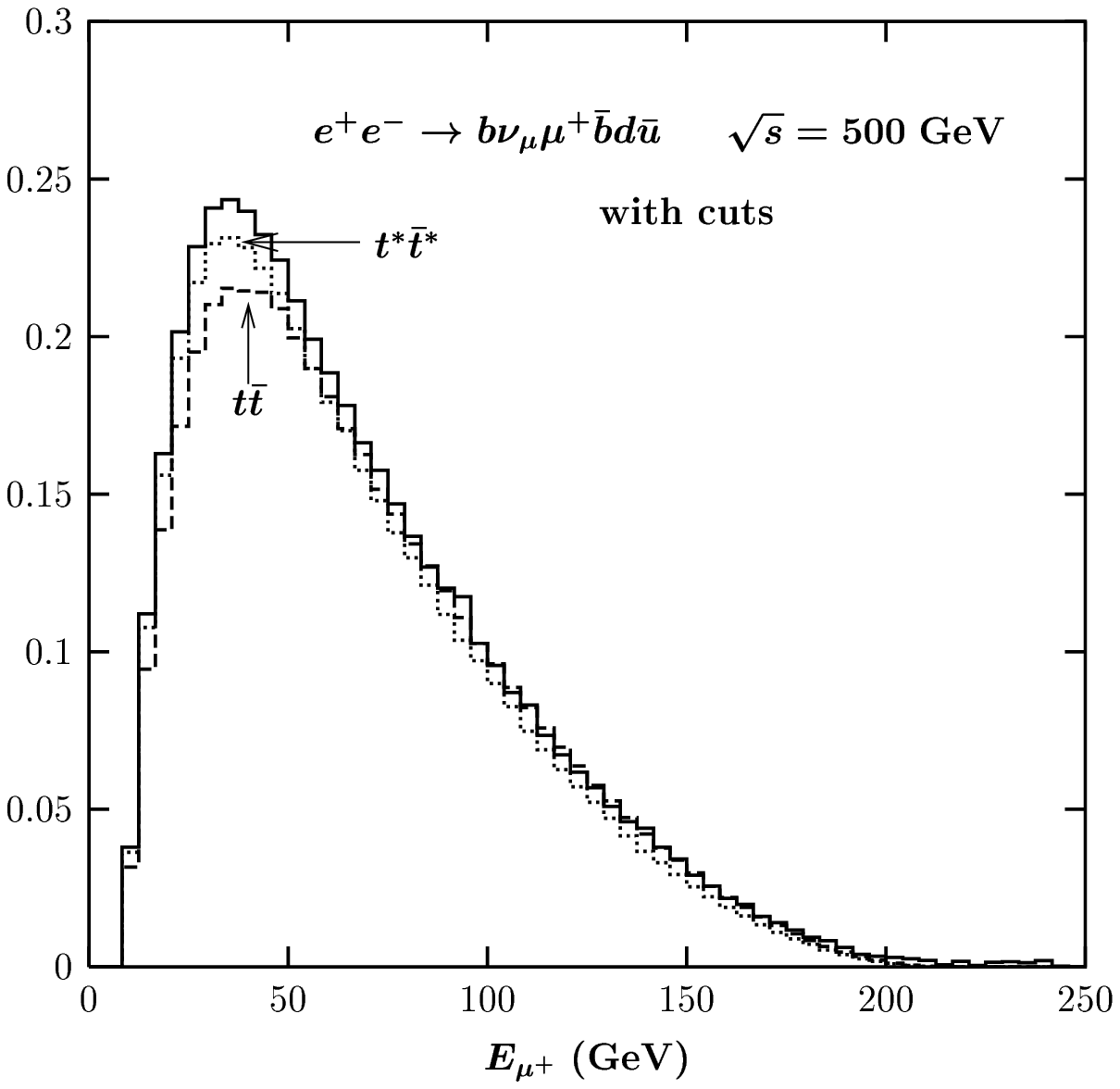}}}
\end{picture}
\end{center}
\vspace*{3.5cm}
\caption{Energy distributions of a $\mu^+$ at $\sqrt{s}=500$ GeV without cuts
and with cuts.}
\end{figure}

\begin{figure}[ht]
\label{fig5}
\begin{picture}(35,35)(-60,-75)
\rput(5.3,-6){\scalebox{0.7 0.7}{\epsfbox{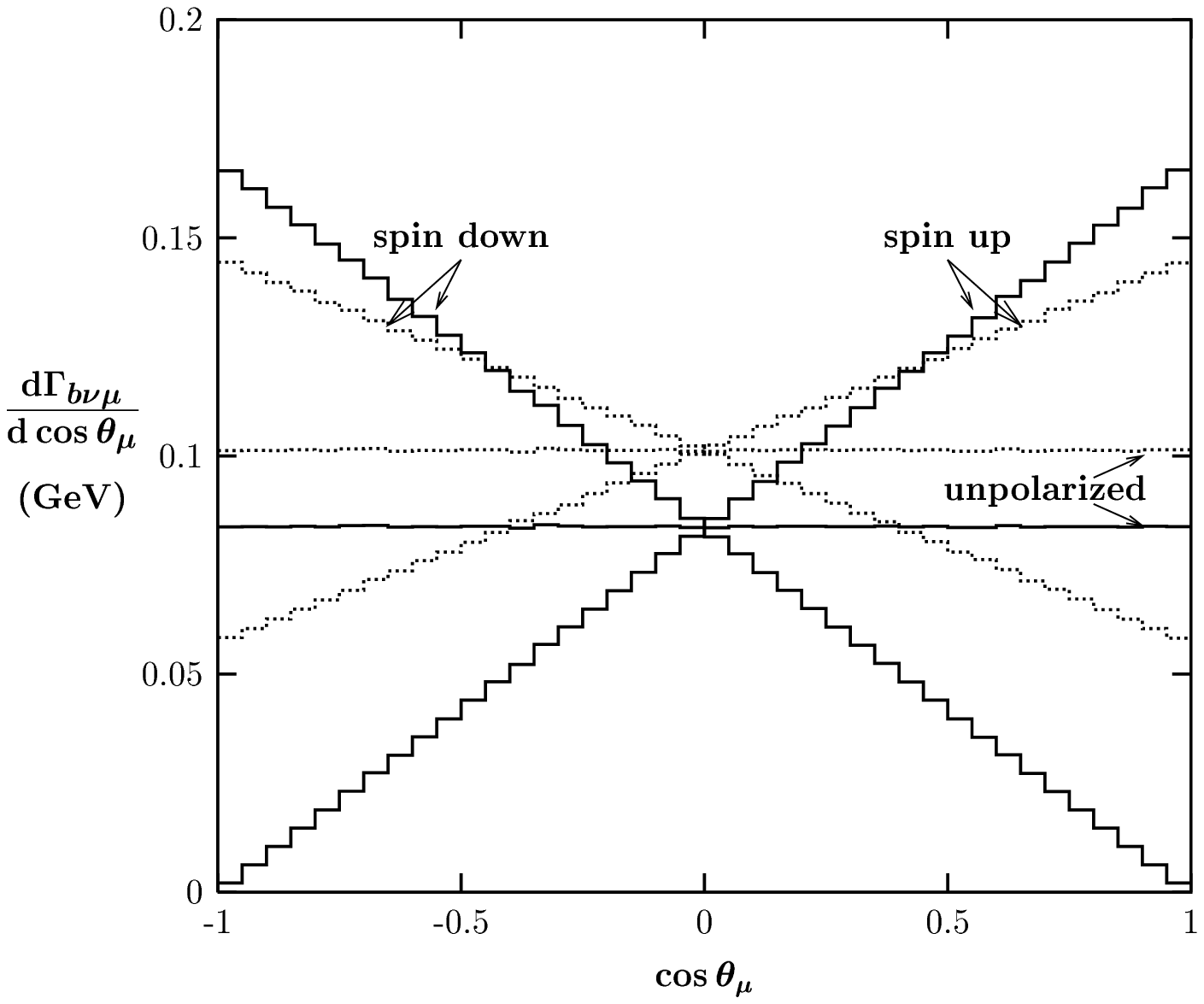}}}
\end{picture}
\vspace*{7. cm}
\caption{Angular distributions of a top quark decay products in the top 
          rest frame.}
\end{figure}

By comparing the plots on the left and right hand 
side of Figs.~2--4, one sees that, with cuts (\ref{cuts}), the background 
and off resonance contributions are reduced to a similar extent as
the top pair production signal itself. Of course, this somewhat
naive analysis does not preclude existence of a set of cuts that 
would keep intensity of the signal and reduced the background at the same 
time, but finding this suitable set need not be a completely simple task 
at all. 

\begin{figure}[ht]
\label{fig6}
\begin{picture}(35,35)(-60,-75)
\rput(5.3,-6){\scalebox{0.7 0.7}{\epsfbox{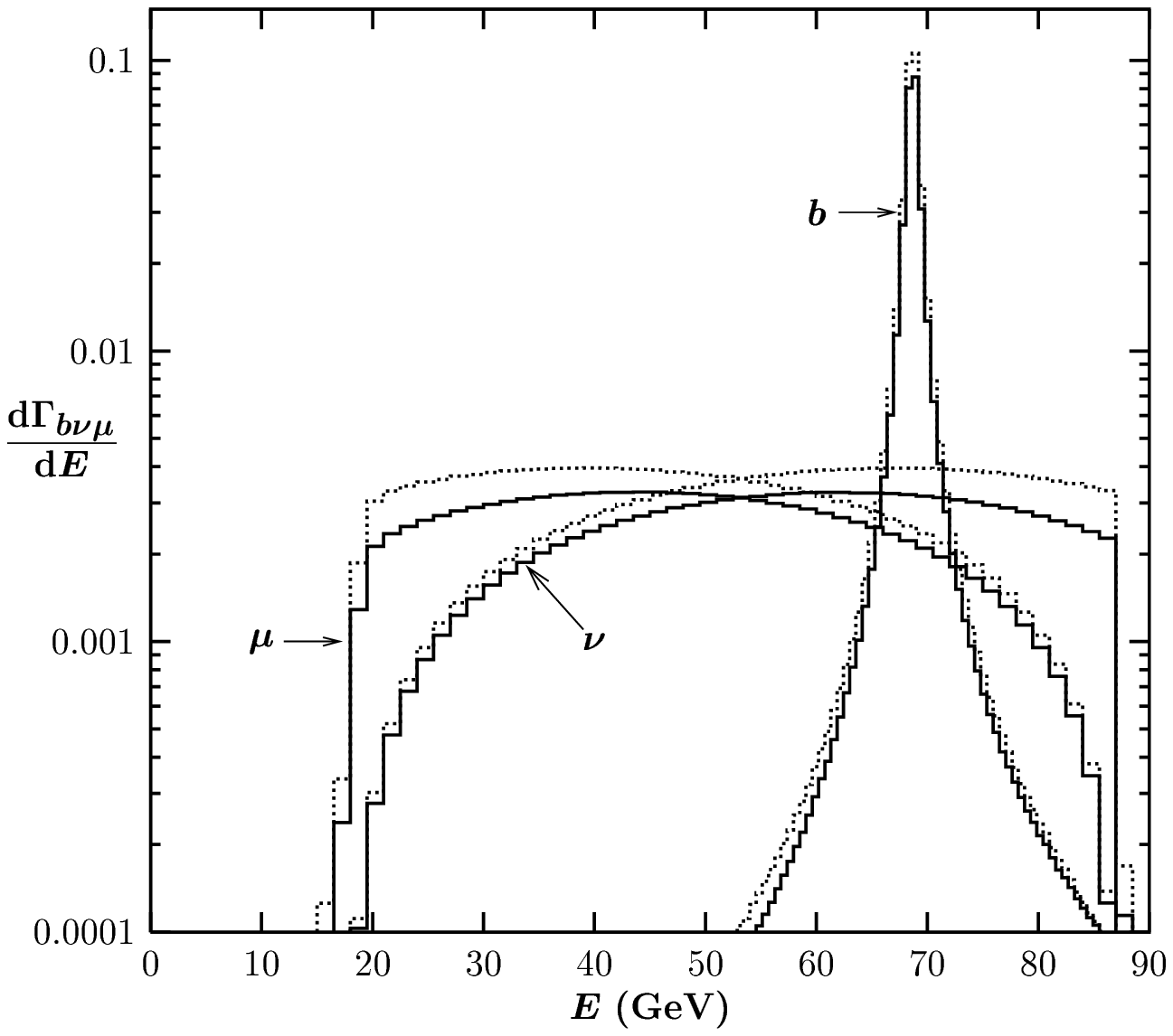}}}
\end{picture}
\vspace*{7. cm}
 \caption{Energy distributions of a top quark decay products in the top 
          rest frame.}
\end{figure}

{\tt eet6f} gives also a possibility of looking at effects of the anomalous 
$Wtb$ coupling defined by Eq.~(\ref{lagr}) that have been implemented in 
the program. Figs.~5 and 6 illustrate how the possible existence
of an anomalous coupling $f_2^+$ may change angular and energy distributions 
of $t$-quark decay products in the rest frame of the top. 
The solid histograms show the SM results, while the
dotted histograms represent results in presence of the anomalous $Wtb$
coupling with $f_1^-=1$, $f_1^+=f_2^-=0$ and $f_2^+=0.1$,
both in Fig.~5 and~6. In Fig.~5, the well known fact from the literature 
that angular distribution of the charged lepton resulting from the decay 
$t \ra b \mu^+ \nu_{\mu}$ is the most efficient analyzer of the top-quar
spin \cite{Jezabek} is recollected. Solid histograms in Fig.~5
clearly show proportionality of the angular distribution 
of $\mu^+$ to $(1 + \cos\theta)$ if the spin of the decaying top-quark 
points in the positive direction of the quantization axis (spin up)
and to $(1 - \cos\theta)$ if the spin of the decaying top-quark 
points in the negative direction of the quantization axis (spin down).


\begin{figure}[ht]
\label{fig7}
\begin{center}
\setlength{\unitlength}{1mm}
\begin{picture}(35,35)(55,-50)
\rput(5.3,-6){\scalebox{0.6 0.6}{\epsfbox{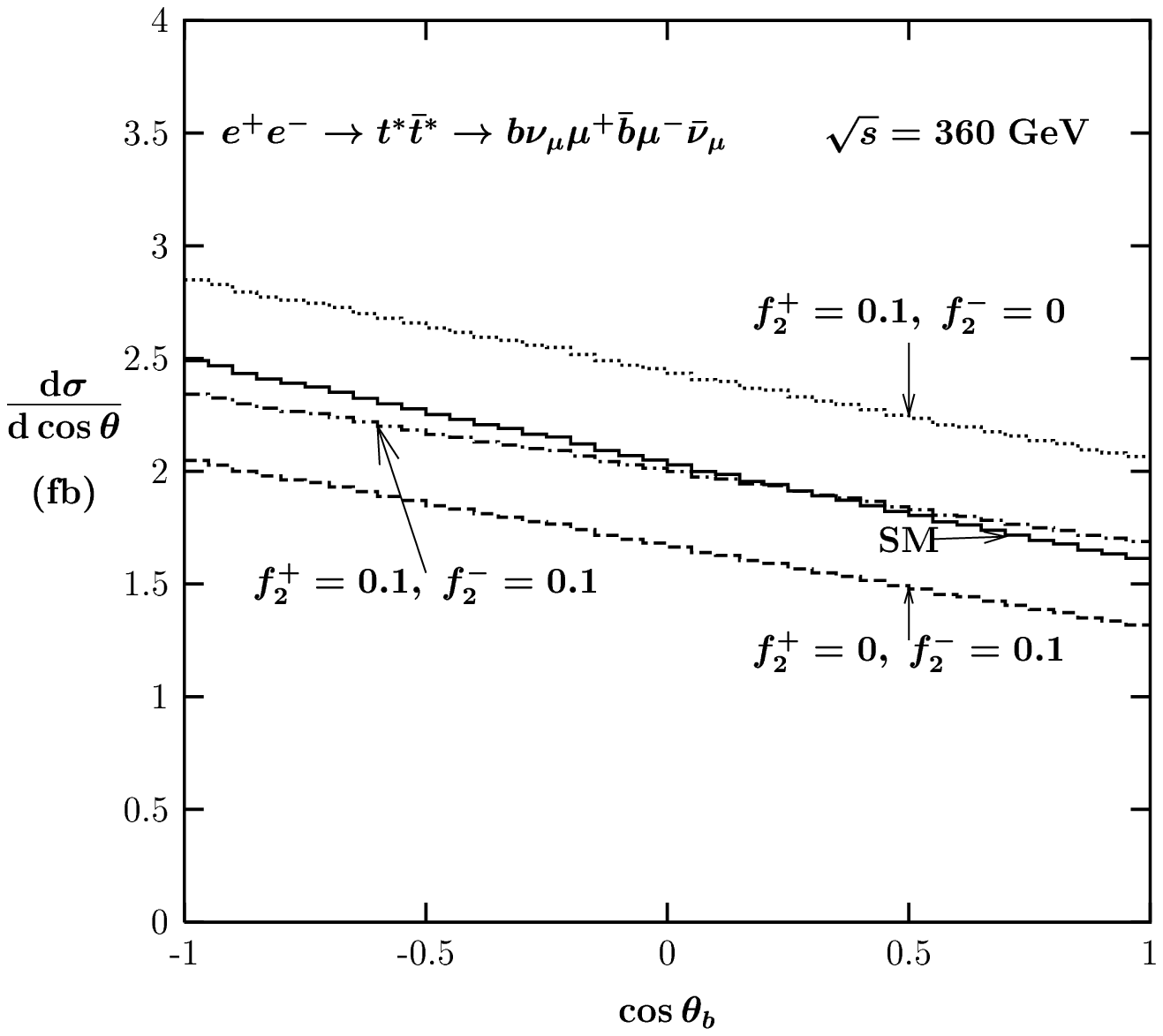}}}
\end{picture}
\begin{picture}(35,35)(15,-50)
\rput(5.3,-6){\scalebox{0.6 0.6}{\epsfbox{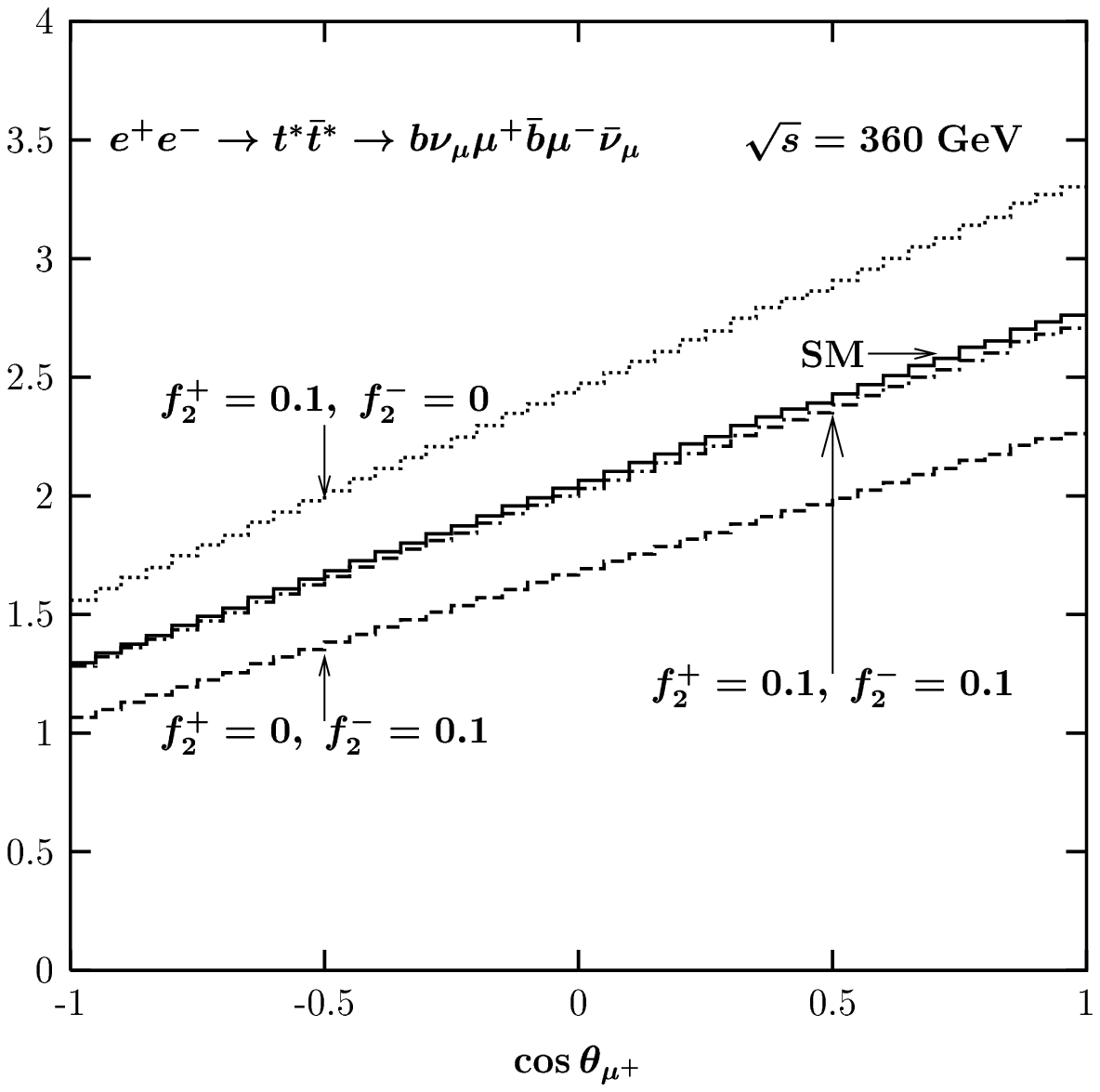}}}
\end{picture}
\end{center}
\vspace*{3.5cm}
\caption{Angular distributions of a $b$-quark (left) and $\mu^+$ (right) 
at $\sqrt{s}=360$ GeV.}
\end{figure}

\begin{figure}[ht]
\label{fig8}
\begin{center}
\setlength{\unitlength}{1mm}
\begin{picture}(35,35)(55,-50)
\rput(5.3,-6){\scalebox{0.6 0.6}{\epsfbox{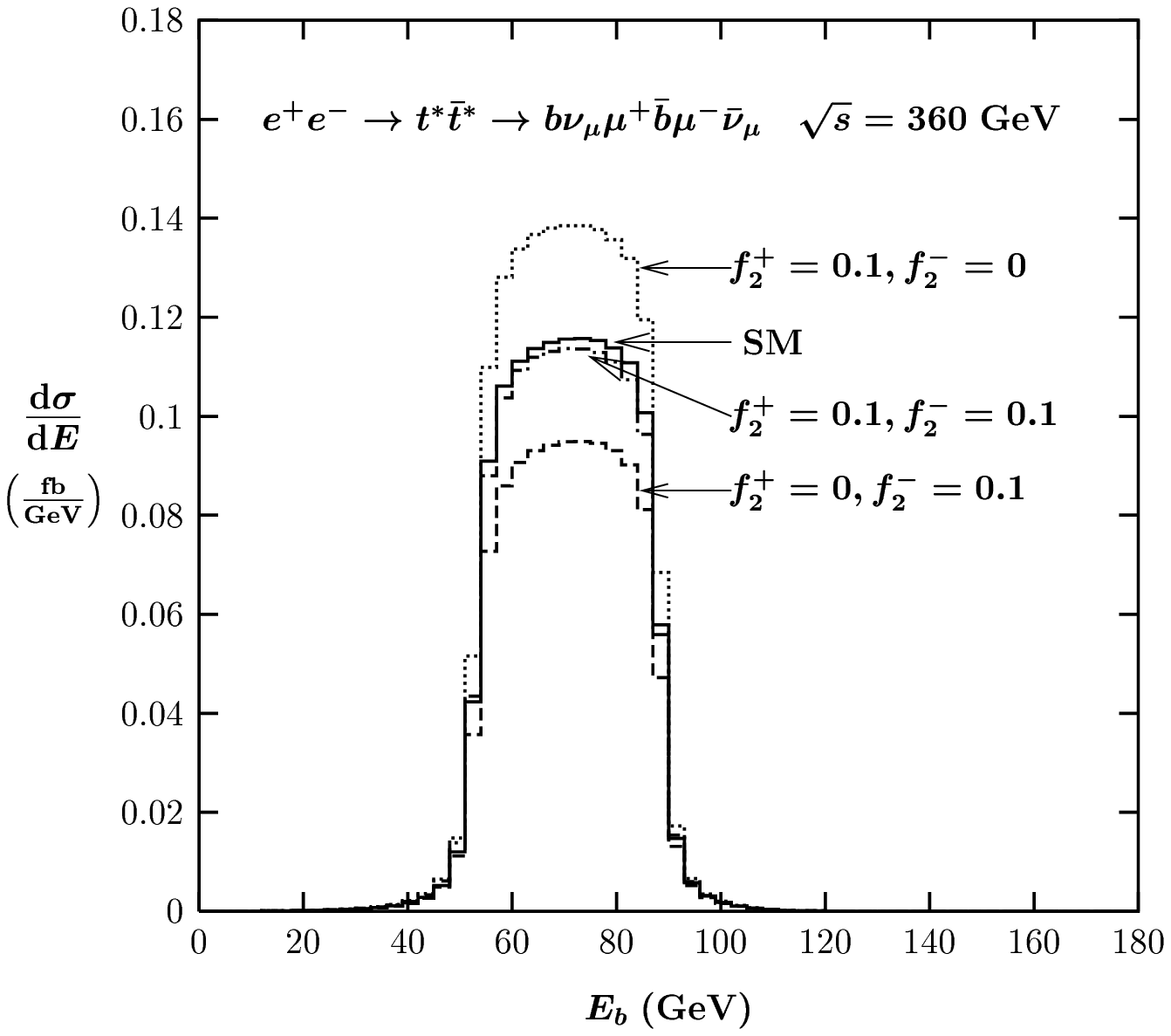}}}
\end{picture}
\begin{picture}(35,35)(15,-50)
\rput(5.3,-6){\scalebox{0.6 0.6}{\epsfbox{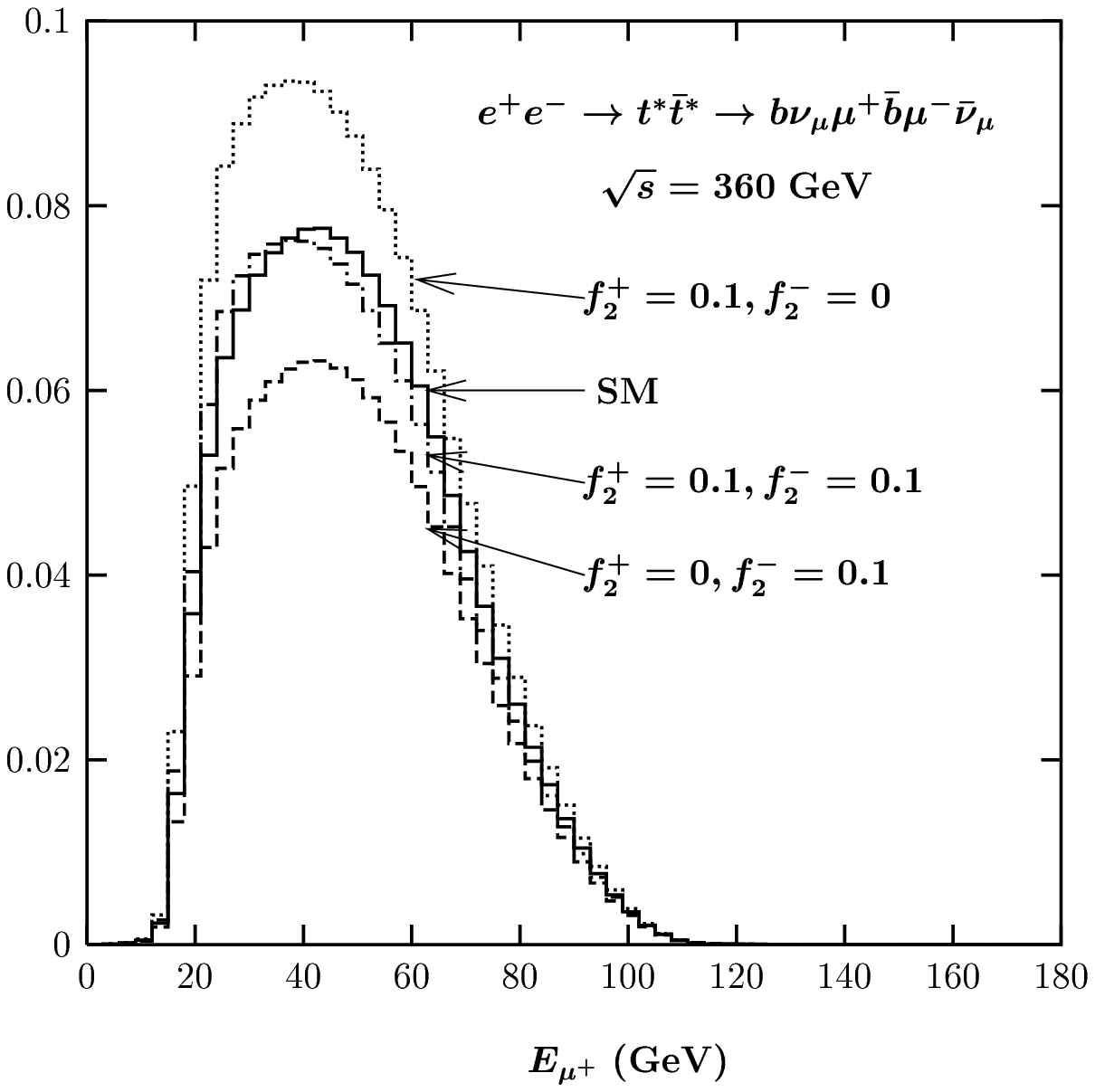}}}
\end{picture}
\end{center}
\vspace*{3.5cm}
\caption{Energy distributions of a $b$-quark (left) and $\mu^+$ (right) 
at $\sqrt{s}=360$ GeV.}
\end{figure}

The angular and energy distributions of a $b$-quark and $\mu^+$ resulting
from the decay of $t$-quark produced in $\epm$ annihilation at 
$\sqrt{s}=360$ GeV are plotted in Figs.~7 and 8. Again the solid histograms
show SM results while the dotted, dashed and dashed-dotted histograms 
represent results in
presence of the anomalous $Wtb$ coupling. One sees that the flat histograms
of Fig.~5 corresponding to decay of the unpolarized top quark have
developed a slope due to the Lorentz boost from the rest frame of 
$t$-quark  to the c.m.s. of $\epm$.  Similarly, how the Lorentz boost 
changes the energy distributions of the top decay 
products is illustrated in Fig.~8. It is interesting to note, how the 
relatively big effects
arising if only a single anomalous coupling is nonzero, {\em i.e.} 
$f_2^+=0.1$ and $f_2^-=0$ (dotted histograms), or $f_2^+=0$ and $f_2^-=0.1$
(dashed histograms), are reduced by interference, if
$f_2^+=0.1$ and $f_2^-=0.1$ (dashed-dotted histograms).

\section{SUMMARY AND OUTLOOK}

Top quark pair production and decay into 6 fermions in $\epm$  annihilation 
at c.m.s. energies typical for linear colliders can be theoretically studied 
to the lowest order of SM with a program {\tt eett6f}. A sample of results 
on selected channels of reaction (\ref{eesixf}) have been presented. It has 
been shown that, although the two $\ttbar$ signal 
diagrams dominate total cross sections even at c.m.s. energies much above the 
$\ttbar$ threshold, the effects related to off-mass-shell production of the
$\ttbar$-pair and the off resonance background may be relevant for the 
analysis of future precision data. It has been illustrated that
these effects are not reduced by imposing typical cuts given by 
(\ref{cuts}). Some extensions of the SM have been implemented in the program 
and results illustrating effects of the nonstandard $Wtb$ coupling 
Eq.~(\ref{lagr}) have been shown.

Work towards completing implementation of all the channels of
(\ref{eesixf}) possible in the SM is underway. In order to match
high precision of future experiments it is also required to implement
higher order effects. As it is not feasible to calculate 
radiative corrections to the full set of Feynman diagrams that contribute
to any specific channel of (\ref{eesixf}), it would
be desirable to include higher order effects at least to the two signal 
diagrams. This approach would be justified by the fact that the total cross 
section of reactions (\ref{eesixf}) is dominated
by the doubly resonant signal. Work in this direction is planned 
in collaboration with the Zeuthen--Bielefeld group \cite{BZ}.

\end{document}